\documentclass[aps,prl,floatfix,twocolumn,superscriptaddress]{revtex4-2}
\usepackage{graphicx} 
\usepackage{stdclsdv} 
\usepackage{array}  
\usepackage{amsmath}
\usepackage[separate-uncertainty=true]{siunitx}
\usepackage{amsfonts}
\usepackage{amssymb}
\usepackage{xcolor}
\usepackage[normalem]{ulem}

\usepackage{hyperref}
\usepackage{xcolor}
\hypersetup{
    colorlinks,
    linkcolor={red!50!black},
    citecolor={blue!50!black},
    urlcolor={blue!80!black}
}

\usepackage{titlesec} % needed package
\titleformat{\paragraph}[runin]
{\normalfont\normalsize\bfseries}{\theparagraph}{1em}{}
\titlespacing*{\paragraph}{0pt}{1ex plus 1ex minus 0.2ex}{\baselineskip} %bfeseris for bold and itshape for italic

\newcommand{\Eq}[1]{Eq.~(\ref{#1})}

\begin{document}

\title{Inferring geometrical dynamics of cell nucleus translocation }

\author{Sirine Amiri}
\thanks{Equal contribution.}
\affiliation{Laboratoire de Physique de l’ENS, École Normale Supérieure, PSL University, CNRS, Sorbonne Université, Université de Paris, Paris, France}

\author{Yirui Zhang}
\thanks{Equal contribution.}
\affiliation{Aix Marseille Univ, CNRS, CINAM, Turing Center for Living Systems, Marseille, France}

\author{Andonis Gerardos}
\affiliation{Aix Marseille Univ, CNRS, CINAM, Turing Center for Living Systems, Marseille, France}

\author{C\'ecile Sykes}
\email{cecile.sykes@phys.ens.fr}
\affiliation{Laboratoire de Physique de l’ENS, École Normale Supérieure, PSL University, CNRS, Sorbonne Université, Université de Paris, Paris, France}
\affiliation{Institut Curie, Paris, France}

\author{Pierre Ronceray}
\email{pierre.ronceray@univ-amu.fr}
\affiliation{Aix Marseille Univ, CNRS, CINAM, Turing Center for Living Systems, Marseille, France}

\begin{abstract}
    The ability of eukaryotic cells to squeeze through constrictions is limited by the stiffness of their large and rigid nucleus. However, migrating cells are often able to overcome this limitation and pass through constrictions much smaller than their nucleus, a mechanism that is not yet understood. This is what we address here through a data-driven approach using microfluidic devices where cells migrate through controlled narrow spaces of sizes comparable to the ones encountered in physiological situations. Stochastic Force Inference is applied to experimental nuclear trajectories and nuclear shape descriptors, resulting in equations that effectively describe this phenomenon  of \emph{nuclear translocation}. By employing a model where the channel geometry is an explicit parameter and by training it over experimental data with different sizes of constrictions, we ensure that the resulting equations are predictive to other geometries. Altogether, the approach developed here paves the way for a mechanistic and quantitative description of dynamical cell complexity during its motility.    
\end{abstract}

\maketitle

\section{Introduction}
 
The cell nucleus, three-to-four times stiffer than the cytoskeleton and twice as viscous, has traditionally been regarded as a mechanically passive compartment housing genetic information \cite{guilak_viscoelastic_2000}. It is now established that in physiological conditions, the nucleus can experience large mechanical stresses that impact its shape and internal organisation, affecting, for example, gene transcription~\cite{uhler_regulation_2017}.
In particular, when cells migrate through complex environments, the nucleus happens to experience large deformations, for instance when passing through tight constrictions~\cite{davidson_nuclear_2014,wolf_physical_2013,patteson_vimentin_2019,estabrook_calculation_2021}. How these large deformations affect nucleus functioning and feed back into the behavior of the cell remain open questions. In fact, the overwhelming majority of cell migration studies focus on experiments on flat surfaces that were crucial to decipher the detailed mechanisms of cell motility~\cite{verkhovsky_self-polarization_1999,maiuri_actin_2015,keren_mechanism_2008}. However, the nucleus is only weakly altered in such experiments, which thus cannot be informative on the role of nuclear mechanics on cell motility, its passive mechanical resistance to deformation, and also the mechanosensory pathways through which these deformations feed back and actuate the cell behavior~\cite{jaalouk_mechanotransduction_2009,maurer_driving_2019,uhler_regulation_2017, venturini_nucleus_2020, lomakin_nucleus_2020}. 

Addressing this problem through \emph{in vivo} experimental observations of cell migration in tightly constraining environments such as the extracellular matrix and epithelial tissues  represents a tremendous challenge. Indeed, one would have to disentangle the complexity of the environment from that of the migrating cell. For this reason, here we study an \emph{in vitro} system of cells migrating in a microfabricated device that imposes three-dimensional mechanical constraints on spontaneously migrating eukaryotic cells~\cite{davidson_design_2015, gundersen_assembly_2018}. We therefore focus on the influence of the geometry on squeezed cell migration. Specifically, cells migrate in an array of pillars designed to impose constrictions of controlled size, which incur large deformations of the nucleus. Remarkably, we find that cells with a nucleus of diameter $\sim12$\unit{\micro\meter} in their rest state are able to spontaneously migrate through constrictions as tight as $2$\unit{\micro\meter}. We refer to this process as \emph{nuclear translocation}, in analogy with polymer translocation where a large macromolecule can pass through tight pores. Using bright-field and multichannel fluorescent imaging, we are able to track the trajectories of individual nuclei going through these constrictions. However, the analysis of the resulting trajectories poses multiple challenges due to their complexity, inherent stochasticity, and the limited amount of data: how does one extract quantitative models and mechanistic insights from such trajectories?

To tackle this challenge, we employ here a data-driven approach to learn dynamical models directly from experimental nucleus trajectories. This contrasts with more traditional model-based approaches that postulate a model form and fit its parameters through the use of aggregate observables such as correlation functions: here we let the model emerge from the data, and the parameters are optimized directly on the entire data set. Such approaches have recently received a lot of attention, in particular due to the development of methods adapted to data-driven inference of deterministic dynamical models such as ordinary and partial differential equations~\cite{brunton_discovering_2016,champion_data-driven_2019}. These methods are well adapted for large-scale datasets such as tissue dynamics~\cite{romeo_learning_2021,schmitt_zyxin_2023}. Importantly however, single-cell dynamics are not deterministic: the inner complexity of these objects, coupled to the reliance to feedback pathways involving small numbers of signalling molecules, results in apparently erratic dynamics which is better captured by stochastic differential equations (SDEs)~\cite{kloeden_numerical_2010}. Data-driven approaches have been used to capture the dynamics of freely migrating cells~\cite{selmeczi_cell_2005,selmeczi_cell_2008,li_dicty_2011}, revealing a persistent random walk behavior. They have been used to quantify the dynamics of non-constraining confined cell migration~\cite{bruckner_learning_2021, bruckner_geometry_2022} and, recently, for constraining cell migration in an elastic environment~\cite{stoberl_nuclear_2023}. Newly introduced inference methods for SDEs~\cite{frishman_learning_2020,bruckner_inferring_2020} have made it possible to efficiently learn such dynamics and have resulted in insights in cell-cell interactions during confined migration that would not have been possible with pre-existing methods~\cite{bruckner_learning_2021}. However, to our knowledge, such methods have not been applied to cell migration with mechanical constraints that lead to large deformations of the nucleus. To this aim, we define and measure quantitative descriptors of the cell shape and state, then use \emph{Stochastic Force Inference} (SFI)~\cite{frishman_learning_2020} to construct a model that captures the dynamics of these shape descriptors. By including the constriction shape as an explicit input of the model, we are able to extrapolate the model to other geometries that could be used for further experimental design.

\section{Results}

\begin{figure}[ht!]
        \includegraphics[width=\columnwidth]{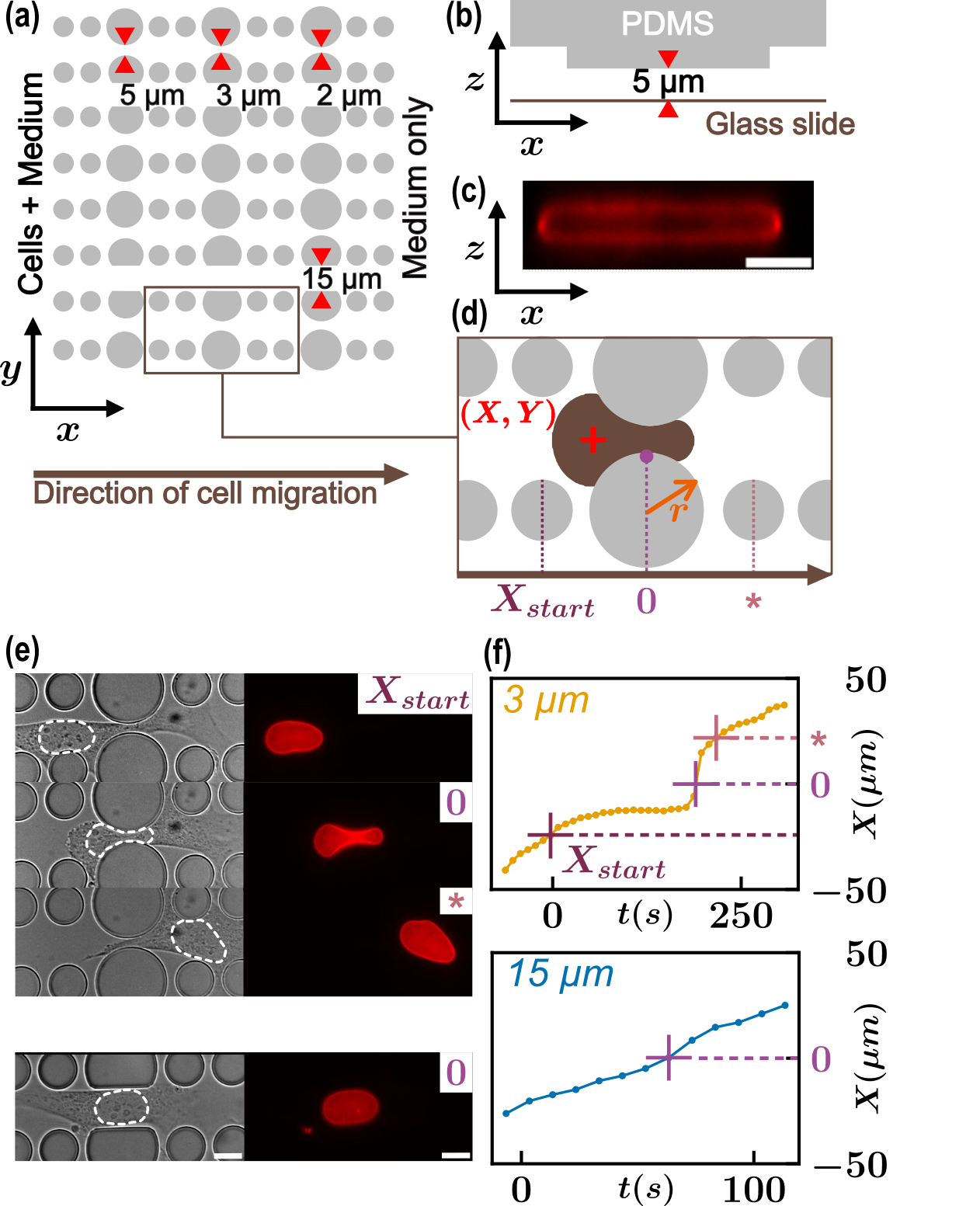}
        \caption{\textbf{\textit{CRISPR engineered MEFs are migrating in a microfluidic device made of constrictions.}} \textbf{(a)} Top view of the pattern used in the microfluidic device. It is composed of PDMS pillars of several widths in order to make three types of constrictions : 5, 3 and \SI{2}{\micro\meter} wide and a control channel of \SI{15}{\micro\meter}. \textbf{(b)} Side view of the microfluidic device. Height of the pillars is 5µm.\textbf{(c)} Side view ($x,z$) of lamin A/C signal in mCherry of an engineered MEF in the middle of a constriction. \textbf{(d)} Representation of the origin points used for a nucleus (\textit{in brown}) trajectory. ($X,Y$) is position of the nucleus (\textit{red star}). \textbf{(e)} Epifluorescence images of an engineered MEF crossing a \SI{3}{\micro\meter} constriction (\textit{top}) and a \SI{15}{\micro\meter} large channel (\textit{bottom}). Left is the transmission signal and right mCherry signal (for Lamin A/C). \textbf{(f)} Examples of trajectories of one cell through a \SI{3}{\micro\meter} constriction (\textit{top}) and one cell through \SI{15}{\micro\meter} large channel(\textit{bottom}). Annotated points corresponds to specific positions illustrated in \textbf{(d)}. Scale bars : \SI{10}{\micro\meter}.} 
        \label{fig:Figure1}
    \end{figure}

\paragraph{Confined cell migration experiments}
We use a CRISPR-modified Mouse Embryonic Fibroblasts (MEFs) cell line that expresses nesprin-2 giant with a green fluorescent protein (GFP) sequence and lamin A/C with a red fluorescent protein (mCherry) sequence \cite{davidson_nesprin2_2020}. The lamin biopolymer shell that lies right underneath the nuclear envelope is linked to the cytoskeleton through the LINC complex, which includes nesprins~\cite{vahabikashi_nuclear_2022, gruenbaum_lamins_2015,tapley_connecting_2013,crisp_coupling_2006,mellad_nesprins_2011}. Cells migrate through microfluidic devices that consist of a series of \SI{5}{\micro\meter} high pillar structures providing three sizes of constrictions (5, 3 and \SI{2}{\micro\meter}) and larger channels (\SI{15}{\micro\meter}) (Fig.\ref{fig:Figure1}a, b). Such migration devices are obtained by covalent assembly of a 3D-imprinted block of polydimethylsiloxane (PDMS) with a glass coverslip~\cite{gundersen_assembly_2018}. Cells are placed on one side of the device with culture medium, before the pillars (See Appendix A1-4). They exhibit global motion (on the $x$-axis) towards the other side, empty of cells but filled with culture medium (Fig.\ref{fig:Figure1}a). The apparent width of MEF cell nuclei (on the $y$-axis) is $12\pm$\SI{2}{\micro\meter} outside of constrictions (See Appendix B1). It is, therefore, larger than constriction sizes and smaller than the large channel of \SI{15}{\micro\meter}. Note that in all conditions, nuclei shapes are mostly cylindrical (on the $z$-axis), touching the bottom and the ceiling of the migration device (Fig.\ref{fig:Figure1}c). We confirm previous observations \cite{davidson_nesprin2_2020} that during nucleus translocation through a constriction, nesprin signal intensity increases at the front of the nucleus while lamin signal does not (See Appendix B2).

\paragraph{Extracting cell nucleus trajectories}
 We observe the movement of cell nuclei in the horizontal ($x,y$) plane when cells migrate between vertical $z$-oriented pillars (Fig.\ref{fig:Figure1}a,b). Nuclei are deformed when they translocate through narrow constrictions \cite{davidson_design_2015}. A constriction is defined by two facing big pillars of radius $r$ (Fig.\ref{fig:Figure1}d). For each constriction, we define the spatial origin ($x=0,y=0$) at the center of the constriction. The position of a nucleus is defined by its surface barycenter ($X,Y$) detected through the lamin signal (Fig.\ref{fig:Figure1}d, right and Appendix A5-6). An image is taken every \SI{10}{\min}. This time interval was optimized to limit fluorescence bleaching and cell phototoxicity.

For each nucleus trajectory, we define $X_{\text{start}}$ as the middle of the small pillar that precedes the specific constriction (Fig.\ref{fig:Figure1}d). The same definition is adapted to the 15\unit{\micro\meter} channels with a truncated-disk pillar defining the "constriction". The start of a trajectory (at time $t=0$ min) is defined either by $X = X_{\text{start}}$ or by its interpolated value using a constant nuclear speed between the two available positions closest to $X_{\text{start}}$. The end of a trajectory is determined by the earliest of i) the end of the overall acquisition, ii) the start of a new trajectory in a new constriction, and iii) half an hour before the cell starts to divide or die. We exclude any trajectory corresponding to cells undergoing adherent cell-cell contact for more than an hour to exclusively address here the migration of individual cells. Examples of recorded images of a nucleus translocating through a \SI{3}{\micro\meter} constriction and a nucleus migrating through a large \SI{15}{\micro\meter} channel are given in Fig.\ref{fig:Figure1}e. The corresponding trajectories and origin points are displayed in Fig.\ref{fig:Figure1}f. We do not observe nuclear rupture during this deformation, contrarily to other mechanical studies of cell nuclei~\cite{denais_nuclear_2016, lomakin_nucleus_2020, pfeifer_gaussian_2022}.

\begin{figure}[tb]
\includegraphics[width=0.8\columnwidth]{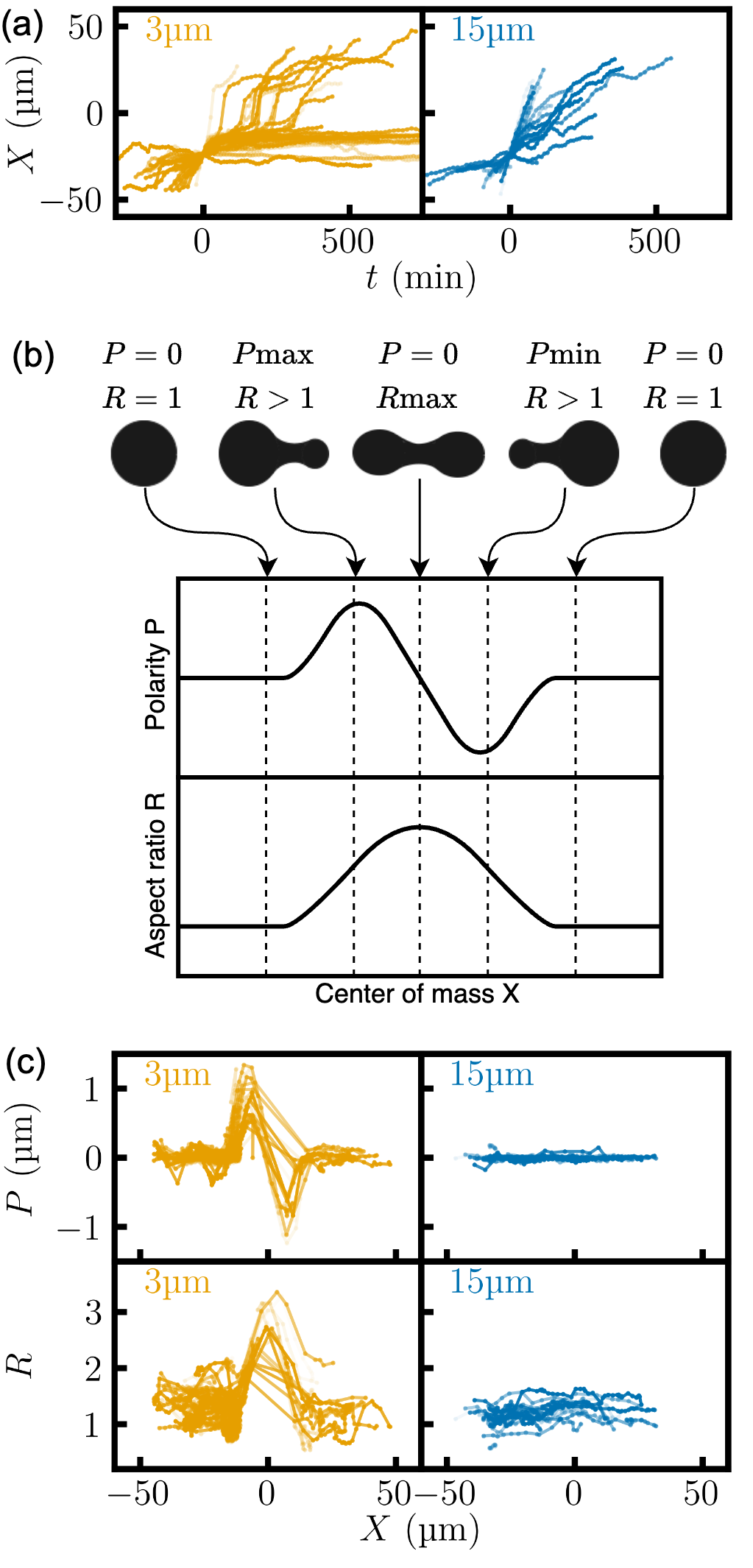}
\caption{\textbf{Experimental trajectories of cell nucleus translocation}. \textbf{(a)} Time series of nucleus position $X$ during translocation. \textbf{(b)} Schematic of characteristic nuclei shapes and their corresponding position during the translocation. \textbf{(c)} Protrusion vector $P$ and aspect ratio $R$ change along the translocation process. In \textbf{(a)} and \textbf{(c)}, results from $3$µm and $15$µm constraints are shown for comparison.}
\label{fig:Figure2}
\end{figure}

A typical nucleus trajectory $X(t)$ through a \SI{3}{\micro\meter} constriction has a sigmoid-like shape, with a plateau soon after $X_{\text{start}}$ when the nucleus reaches the entrance of the constriction, and a sharp acceleration when it manages to pass through the center of the constriction at $X = 0$, followed by an unconstrained motion (Fig.\ref{fig:Figure1}f, top). A nucleus trajectory in a large \SI{15}{\micro\meter} channel displays a smooth movement (Fig.\ref{fig:Figure1}f, bottom) at almost constant velocity. However a fraction of cells do not translocate before the end of the trajectory recording. They are nevertheless included in our data set to avoid any statistical bias in the analysis. Overall, nucleus trajectories show some variability, both in the duration of the plateau and in the velocity of free migration, as can be seen in Fig.\ref{fig:Figure2}a. 

\paragraph{Modeling the stochastic nature of nucleus translocation}
The observed variability between different trajectories in the same constriction condition (either the \SI{3}{\micro\meter} or the \SI{15}{\micro\meter} wide constriction) and the fluctuations of velocity during a single trajectory reflect the internal complexity of the cells. 
This effective stochasticity is often modeled through a noise term by using stochastic differential equations (SDEs). A SDE describing the cell nucleus motility is typically of the form \cite{bruckner_learning_2023} :
\begin{equation}
\dot{X}(t) = \underbrace{\Pi + f_{ext}(X)}_{\textrm{deterministic}} + \underbrace{\sqrt{2D_{X}}\cdot \eta_{X}(t)}_{\textrm{stochastic}}, 
\label{eqn:CM-x}
\end{equation}
which consists of two deterministic terms ($\Pi$ and $f_{ext}$) reflecting the slow, predictable aspects of the dynamics and a stochastic noise that models the coupling of the observed position with fast, unobserved degrees of freedom. More specifically, in the deterministic contribution of nucleus dynamics, $\Pi$ is a driving term, also called \emph{polarity} of the cell, and captures the asymmetry in the internal organization of the cell that generates the motility \cite{drubin_origins_1996}. The other deterministic term, $f_{ext}$, represents the direct effect of the environment on cell nucleus dynamics. The noise term $\sqrt{2D_{X}} \cdot \eta_{X}$ has an amplitude characterized by its diffusion coefficient $D_{X}$, which we assume here to be state independent, and $\eta_{X}$ the noise, which for simplicity we assume to be white and Gaussian, therefore, $\langle \eta_{X}(t) \rangle = 0$ and $\langle \eta_{X}(t) \eta_{X}(t')\rangle = \delta(t - t')$. 

The polarity $\Pi$ itself is dynamical, and its dynamics describe the way cells sense their environment and actuate their self-propulsion accordingly. The dynamics of $\Pi$ follows an SDE of the form
\begin{equation}
\dot{\Pi}(X, t) = f_{\Pi}(X, \Pi) + \sqrt{2D_{\Pi}}\cdot \eta_{\Pi}(t), 
\label{eqn:overdamp-P}
\end{equation}
The drift term $f_{\Pi}(X, \Pi)$ encodes the internal dynamics of $\Pi$ as well as the feedback of the nucleus polarity to the external environment.  
Note that there are thus two ways the environment affects the dynamics: through direct forces on $X$ (term $f_{ext}(X)$ in Eq.\ref{eqn:CM-x}) and through indirect feedback ($f_{\Pi}(X, \Pi)$ in Eq.\ref{eqn:overdamp-P}) -- \emph{i.e.} mechanosensing. Here again, fast internal processes of the cell are modeled as a Gaussian white noise $\sqrt{2D_{\Pi}}\cdot \eta_{\Pi}(t)$ with diffusion coefficient $D_{\Pi}$, which determines, for instance, the persistence length of the free motion of the cell~\cite{li_dicty_2011}.

The class of cell motility models described by Eqs.~\ref{eqn:CM-x} and~\ref{eqn:overdamp-P} is very general and widely used. However, a key challenge to its applicability to experimental data is that the polarity 
 $\Pi$ is not directly measurable, as its molecular definition remains unknown. To bypass this difficulty, previous studies have relied on the use of underdamped dynamics: briefly, such approaches consist of differentiating \Eq{eqn:CM-x} with respect to time, and plugging into \Eq{eqn:overdamp-P} to eliminate $\Pi$, thus resulting in an effectively second-order dynamics for $X$~\cite{bruckner_learning_2023}. This type of \emph{embedding} approach exploits Taken's theorem and is popularly used for deterministic dynamical systems~\cite{crutchfield_equations_1987}. While this approach has been successful in quantifying, for instance, cell-cell interactions from data~\cite{bruckner_learning_2021}, it has several drawbacks. First, second-order inference is considerably more difficult and demanding in terms of data quality and precision than first-order inference~\cite{bruckner_inferring_2020}. Second, one has to neglect the noise on nucleus position in order for this approach to work, which is not always possible. Third, information about the nature of the polarity and its feedback mechanisms is lost in the process. An alternative approach was proposed recently, consisting of a model-driven treatment of data where the polarity is explicitly included as a hidden variable~\cite{bruckner_geometry_2022}, but this requires strong assumptions on the motility mechanisms.

\paragraph{Data-driven modeling from geometric quantities}
We propose here an alternative approach to maintain the overdamped dynamics, which is more physical, and approximate $\Pi$ with available information. 
Indeed, we have access to more than just the nuclear center $X$: using the lamin signal, we can track the precise contour of the nucleus and extract a richer set of geometrical quantities. In particular, when the cell engages into the constriction, the nucleus starts elongating and protruding toward the narrow part of the constriction, as schematized in Fig.\ref{fig:Figure2}b. When exiting the constriction, the protrusion points backwards, and the nucleus progressively recovers is oval shape.

From these observations, we define two variables to characterize nucleus deformations.  First, to account for the geometrical shape change of the nucleus, we define its \emph{protrusion vector} $P=X_{c}-X$, with $X_{c}$ the barycenter of the contour of the nucleus (see detailed expression in Appendix A7). The quantity $P$ gives a measure of how much and in which direction the nucleus boundary protrudes relative to the center of mass. A positive (resp. negative) value of $P$ corresponds to a forward (resp. backward) extension of the nucleus relative to the center of mass (see Fig.\ref{fig:Figure2}b). 
Second, we characterize the relative ($x,y$) stretch by defining the aspect ratio of the cell nucleus $R$ (see detailed expression in Appendix A7). As illustrated in Fig.\ref{fig:Figure2}b, a circular disk corresponds to $R = 1$, whereas $R < 1$ (resp. $R > 1$) corresponds to a horizontal (resp. vertical) ellipsoid.

The quantities $P$ and $R$ describe two different and complementary aspects of nucleus deformation. As illustrated in Fig.\ref{fig:Figure2}b, $P$ does not distinguish a dumbbell shape from a sphere, whereas $R$ does; $R$ cannot distinguish a front protrusion from a back protrusion, whereas $P$ does. Examples of the time series $P$ and $R$ against $X$ when cells go through a $3$\unit{\micro\meter}-constriction (left) or a $15$\unit{\micro\meter}-channel (right) are shown in Fig.\ref{fig:Figure2}c. This shows that the constriction affects significantly $P$ and $R$ compared to the \SI{15}{\micro\meter} large channels. We then make use of these complementary geometrical data to learn a quantitative model for nuclear translocation dynamics.

%========================================
\paragraph{Inferring coupled dynamics of position and geometry}
Our aim is to obtain a data-driven, quantitative, autonomous description of nuclear translocation using the position $X$ and geometric descriptors $P$ and $R$.
More precisely, for each recorded nucleus trajectory in constraints $2, 3, 5$ and \SI{15}{\micro\meter}, we extract three time series $\{X_{t}, P_{t}, R_{t}\}$ at acquisition times $t = 0, \Delta t, 2\Delta t\dots$. These data serve as the input in our inference analysis, from which we aim to extract coupled SDEs capturing the continuous-time dynamics of $(X_t, P_t, R_t)$. We postulate that including the geometric quantities $P_t$ and $R_t$, on top of the nucleus position $X_t$, makes it possible to identify a set of such equations that is both autonomous (\emph{i.e.} that does not couple to the dynamics of other, unobserved quantities, in contrast to the approach of Ref.~\cite{bruckner_geometry_2022}) and physically first-order (\emph{i.e.} that does not introduce emergent inertia as a polarity model, in contrast to most pre-existing literature~\cite{li_dicty_2011,bruckner_stochastic_2019,bruckner_learning_2021,bruckner_learning_2023}).

To achieve this, we analyze the time series using a recently introduced framework, Stochastic Force Inference (SFI)~\cite{frishman_learning_2020}. SFI allows us to reconstruct first-order SDEs from such time series by employing a data-efficient quasi-maximum-likelihood linear regression algorithm. In practice, it consists of approximating the drift term with estimators formed by a linear combination of basis functions. Here, we start from a relatively large basis that we construct based on symmetries and our physical understanding of the quantities we model, and that include a systematic expansion of the geometrical features of the system -- \emph{i.e.} the $x$-dependent constriction width. We then iteratively reduce this basis to select an appropriately minimal model for the dynamical equations we aim to learn. More specifically, our starting model is:
\begin{equation}
\begin{split}
\dot{X}_t
    = &\overbrace{C_{X} + \alpha_{X} P_t + \beta_{X}(R_t -R_t^{-1})}^{\mathrm{internal\ driving}} + f_{X}(X_t, r) \\
    &+ \sqrt{2D_{X}}\cdot\eta_{X}(t),    
\end{split}
\label{eqn:generic-1}
\end{equation}
\begin{equation}
\begin{split}
    \dot{P}_t %&= f_{P}(X, P, R) + \sqrt{2D_{P}}\cdot\eta_{P}(t) \\
    &= \overbrace{C_P + \alpha_{P} P_t + \beta_{P}(R_t-R_t^{-1})}^{\mathrm{internal\ dynamics}} + f_{P}(X_t, r) \\ &+ \sqrt{2D_{P}}\cdot\eta_{P}(t),
\end{split}
\label{eqn:generic-2}
\end{equation}
\begin{equation}
\begin{aligned}
    \dot{R}_t %&= f_{R}(X, P, R) + \sqrt{2D_{R}}\cdot\eta_{R}(t) \\
    &= \overbrace{C_R + \alpha_{R} P_t + \beta_{R} (R_t - {R}_t^{-1})}^{\mathrm{internal\ dynamics}} +  f_{R}(X_t, r) \\&+ \sqrt{2D_{R}}\cdot\eta_{R}(t). 
\end{aligned}
\label{eqn:generic-3}
\end{equation}
Eq.\ref{eqn:generic-1} connects to the general form presented in Eq.\ref{eqn:CM-x} by approximating the polarity $\Pi$ with a linear combination of three terms: $C_{X}$, a constant drift representing the propensity of cells to migrate in the $x$ direction, physically motivated by the imbalance in cell populations between the two sides of the device; $\alpha_{X} P$ which is a vector-like term coupling the direction of motion and protrusion direction; and $\beta_{X}(R -R^{1})$ by which the nucleus shape modulates the self-propulsion velocity around its rest shape $R=1$. The remainder, $f_{X}$, captures the effect of the environment, and thus depends on the position $X_t$ -- we omit, for simplicity, the possibility that it depends on the geometry. Similarly, the dynamics of $P$ is described by Eq.\ref{eqn:generic-2} (resp.~$R$ by Eq.\ref{eqn:generic-3}) with the same decomposition into internal dynamics and external influence, and we use the same basis functions. Note that we use the combination $(R-1/R)$ to reflect the fact that the aspect ratio $R$ is a ratio of lengths which should remain positive at all times, and has average value of $1$ in the absence of external constraints.

In a complex or unknown environment, the drifts $f_{X}$, $f_P$ and $f_R$ representing the influence of the environment on $X, P$ and $R$ would have to be expanded on a generic basis. Here, however, we take advantage of the fact that the geometry of the channel is known to simplify inference and allow for extrapolation of the model to other geometries. Specifically, we include the radius $r$ of the pillars that form the constriction (see Fig.~\ref{fig:Figure1}d) as an explicit parameter of the inference, and construct our basis functions using the channel width $w(X, r)$ as well as the normal to the pillar $\hat{n}(X, r) = (n_{x}(X, r), n_{y}(X, r))$. (See Appendix A8 for a schematic of these quantities and their expressions).
Using $w(X, r), n_{x}(X, r)$ and $n_{y}(X, r)$ as ingredients, we approximate the environmental drift the nucleus experiences and reacts to, $f_{X}(X, r), f_{P}(X, r)$ and $f_{R}(X, r)$ in the constraint formed by pillars of radius $r$. 
Integrating the pillar radius as a control parameter into these functions allows us to infer a single model for the whole experimental data set of different constriction sizes, including the reference case where the channel does not have a constriction. It makes the model more straightforward and easier to interpret and allows us to use the data more efficiently.
As the influence of the pillars on the cell nucleus is expected to increase with decreasing channel width, we expand this geometrical influence in an inverse power series of the channel width in the basis, up to third-order, \emph{i.e.,} $1/w$,  $1/w^{2}$ and  $1/w^{3}$, which we multiply by geometrical quantities $1$, $n_x$ and $n_y$ that capture distinct features of the constriction. The full expression of our initial model is summarized in Appendix A9.

The SFI algorithm provides estimators for the coefficients of the drift field as a  linear combination of these basis functions. The initial model, which consists of the complete set of basis functions (in total $36$), is shown in Appendix A9.  
A challenge to the use of stochastic inference techniques on cell migration data is that the time interval between frames $\Delta t$ is typically of the same order as the typical translocation time, and cannot be easily decreased as more frequent imaging would incur phototoxicity. 
To overcome this problem, we introduce an improvement on the SFI algorithm to accommodate large time steps, which uses a trapezoidal integration scheme that results in lower discretization biases than previous methods (see Appendix A10 for details). 

\begin{figure}[tb]
\includegraphics[width=0.48\textwidth]{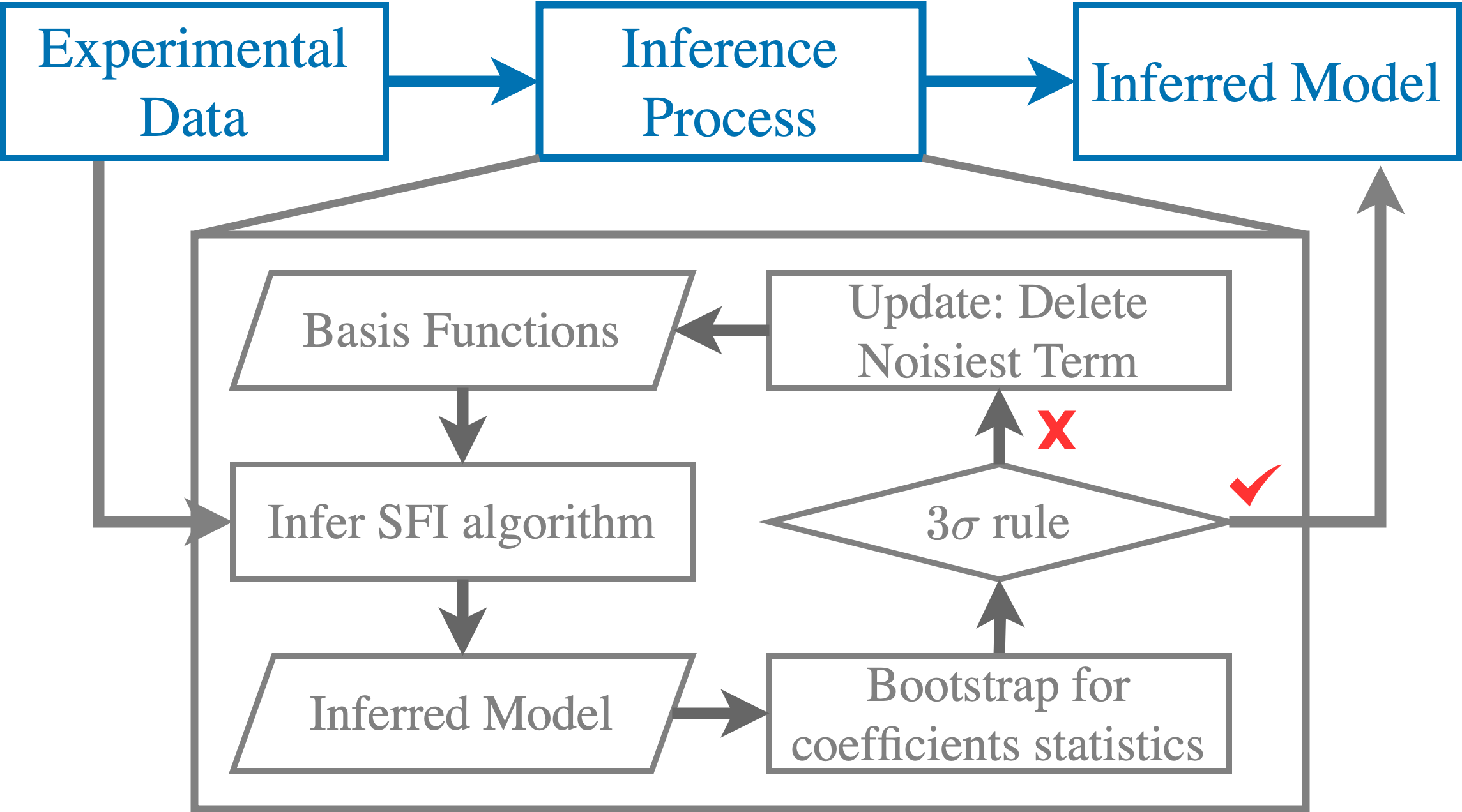}
\caption{\textbf{Schematic of the inference workflow}. To start, write down an initial model -- an overdamped Langevin equation of the problem at hand. Propose the basis functions that form the drift part of the equation and fix any known coefficients.  \textbf{Step 1. Infer}, input the experimental data and the candidate model into the SFI algorithm to obtain the most probable value of the unknown coefficients. \textbf{Step 2. Evaluate}, using bootstrap to obtain the mean and the standard deviation for each coefficient. We evaluate the significance of each coefficient against the $3\sigma$ rule. If one or more coefficients fails this test, then \textbf{Step 3. Update}, update the model by removing the noisiest term. Repeat this process until a final model is reached, where any further elimination would deteriorate the model.}
\label{fig:workflow}
\end{figure}

%========================================
\paragraph{Model Selection algorithm}
%========================================
The learned model consisting of the full set of basis functions is constructed through physically motivated systematic expansion, and as such it is not minimal, which potentially leads to overfitting the data and precludes physical interpretation. To overcome this difficulty and obtain a more interpretable model, we improve this model through a sparsity-enforcing algorithm that consists in iteratively deleting the least statistically significant terms until a threshold significance is reached. This inference workflow, as schematized in Fig.\ref{fig:workflow}, differs from popular sparse learning algorithms that include a penalization based on the values of the coefficients~\cite{brunton_discovering_2016,boninsegna_sparse_2018,callaham_nonlinear_2021}, which would not be appropriate here due to the fact that coefficients have distinct physical dimensions. 

More specifically, this workflow consists of three iterative steps: \textit{infer}, \textit{bootstrap}, and \textit{update}. The first step \emph{infer} uses the SFI algorithm to learn coefficients using the current set of basis functions.
In the second step \emph{bootstrap}, we assess the statistical significance of each inferred coefficient using the bootstrap method, running the inference again on sets of trajectories sampled with replacement and using the standard deviation of the coefficients as a confidence interval (see Appendix A11 for details \cite{tibshirani_introduction_1994}). The significance of each basis function for our model is quantified by their signal-to-noise ratio -- \emph{i.e.} the ratio between the absolute value of the mean of the coefficient and its standard deviation.
If one of these ratios is below a chosen significance threshold of $3$ (corresponding to a $3\sigma$ rule), we move to the third step \emph{update}, where we simplify the model by removing the least significant function from the basis, and iterate the process. The outcome of this process is a final, minimal model where all terms are statistically significant.

\paragraph{Resulting model for nuclear translocation dynamics}

Applying this inference workflow to the whole nuclear translocation data set, we obtain the following model: 
\begin{equation}
    \begin{aligned}
        \dot{X}_t = & C_X + \alpha_{X} P_t + \beta_{X}(R_t -R_t^{-1}) + \dfrac{a_{X}}{w^2(X_t, r)} +\\ & b_{X}\dfrac{n_{y}(X_t, r)}{w^{2}(X_t, r)} + \sqrt{2D_{X}}\cdot\eta_{X}(t),
    \end{aligned}
\label{eqn:final-1}
\end{equation}
\begin{equation}
    \begin{aligned}
        \dot{P}_t = & \alpha_{P} P_t + \beta_{P} (R_t-R^{-1}_t)  + a_{P} \dfrac{1}{w(X_t, r)} + b_{P} \dfrac{1}{w^2(X_t, r)} \\
        &+ c_{P} \dfrac{n_{X}(X_t, r)}{w^2(X_t, r)} + \sqrt{2D_{P}}\cdot\eta_{P}(t),
    \end{aligned}
\label{eqn:final-2}
\end{equation}
\begin{equation}
    \begin{aligned}
        \dot{R}_t = & C_R + \alpha_{R} P_t + \beta_{R} (R_t - R_t^{-1}) +  a_{R}\dfrac{1}{w(X_t, r)} \\
        &+ \sqrt{2D_{R}}\cdot\eta_{R}(t),
    \end{aligned}
\label{eqn:final-3}
\end{equation}
with a total of 14 drift terms. The values and standard deviations of the corresponding coefficients, as well as the inferred diffusion constants, are shown in Table \ref{table: coeffs}.

\begin{table}[b]
\centering
\begin{tabular}{|c||c|c|c|}\hline
 Corr. term&Coeffs.& Value& Unit\\\hline
 $\cdot$&$C_{X}$ & \num{2.6\pm0.5e-1} &\unit{\micro\meter\min^{-1}}
\\
 $P$&$\alpha_{X}$ & \num{-1.7\pm0.2e-1} &\unit{\min^{-1}}
\\
 $R-1/R$&$\beta_{X}$ &\num{6.7\pm0.7e-2} & \unit{\micro\meter\min^{-1}}
\\
 $1/w^2$&$a_{X}$ & \num{-4.7\pm1.1e1} &\unit{\micro\meter^{3}\min^{-1}}
\\
 $n_y/w^2$&$b_{X}$ & \num{4.8\pm1.1e1} & \unit{\micro\meter^{3}\min^{-1}}
\\
 $\eta_X$& $D_{X}$ & \num{5.8\pm0.8e-2} &\unit{\micro\meter^{2}\min{-1}} \\
\hline%------------------------------------------------------------
 $P$&$\alpha_{P}$ & \num{-3.2\pm0.6e-2} &\unit{\min^{-1}}
\\
 $R-1/R$&$\beta_{P}$ & \num{3.0\pm0.7e-3} &\unit{\micro\meter\min^{-1}}
\\
$1/w$ &$a_{P}$ & \num{7.8\pm1.3e-2} &\unit{\micro\meter^{2}\min^{-1}}
\\
$1/w^2$ &$b_{P}$ & \num{-1.0\pm0.2} &\unit{\micro\meter^{3}\min^{-1}}
\\
$n_x/w^2$&$c_{P}$ &\num{-3.1\pm0.5} &\unit{\micro\meter^{3}\min^{-1}}\\
$\eta_P$& $D_{P}$ & \num{8.0\pm1.4e-4}  &\unit{\micro\meter^{2}\min{-1}} \\
 \hline%------------------------------------------------------------
$\cdot$ & $C_{R}$ & \num{-5.7\pm1.2e-3} &\unit{\min^{-1}} 
\\
 $P$&$\alpha_{R}$ & \num{2.2\pm0.3e-2} &\unit{\micro\meter^{-1}\min^{-1}} 
\\
 $R - 1/R$&$\beta_{R}$ & \num{-8.7\pm1.4e-3} & \unit{\min^{-1}} 
\\
 $1/w$&$a_{R}$ & \num{1.1\pm0.2e-1} & \unit{\micro\meter\min^{-1}}
\\
 $\eta_R$&$D_{R}$ & \num{8.6\pm1.8e-4} & \unit{\min^{-1}} \\
 \hline%------------------------------------------------------------
\end{tabular}
\caption{Inferred coefficients for the minimal model, with corresponding terms in Eqs.~\ref{eqn:final-1}-\ref{eqn:final-3}. The confidence intervals correspond to the standard deviation obtained through bootstrapping.} 
\label{table: coeffs}
\end{table}

\begin{figure}[tb]
\includegraphics[scale=0.9]{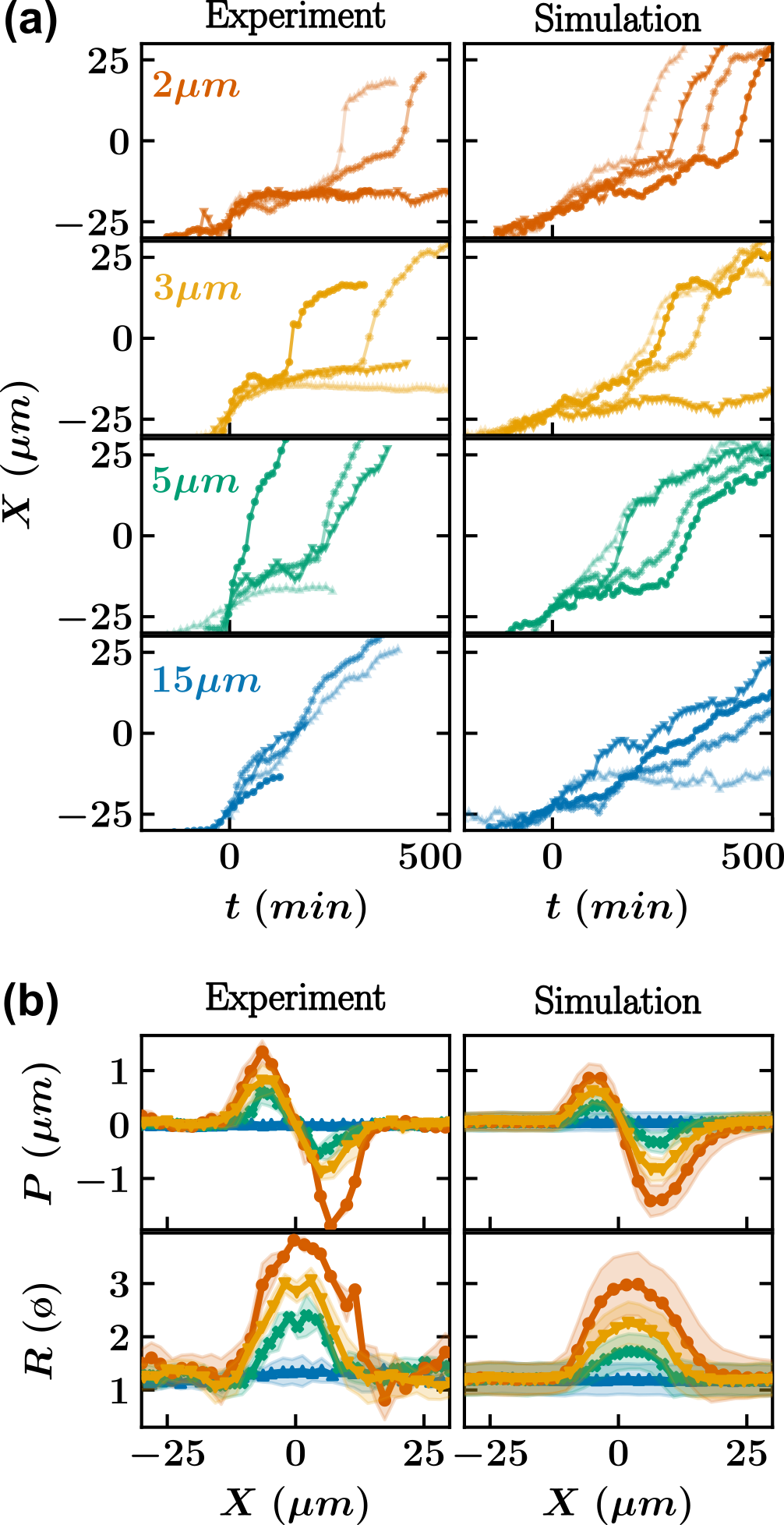}
\caption{\textbf{Simulation results of the reduced model.} \textbf{(a)} Comparison of four representative time series of cell nuclei position $X(t)$ from the experimental data and simulation. \textbf{(b)} Experiment and simulation comparison of the averaged trajectories of boundary polarity against nuclei position $P(X)$ and nuclei aspect ratio against nuclei position $R(X)$.}
\label{fig:Figure4}
\end{figure}

 A representative selection of trajectories $X(t)$ from the experiment and simulation is given in Fig.\ref{fig:Figure4}a. Averaged trajectories $(P, X)$ and $(R, X)$ are given for experiments and simulations in Fig.\ref{fig:Figure4}b ($N=\num{1000}$ simulated trajectories for the averaged quantities). The simulation agrees with the experiments on the general shape of the curves $P(X)$ and $R(X)$, as well as the starting and ending points of the deformation. Position-binned curves $P(X)$ and $R(X)$ at different constraints can be differentiated: the deformation and protrusion increase significantly as the constraint becomes smaller. 
 
Physically, the fact that $C_X>0$ indicates an average propensity of cells to migrate towards the nutrient-rich region. Interestingly, we find that $\alpha_X<0$ and $\beta_X>0$: when entering the constricted region, the cell slows down as the nucleus first protrudes, then accelerates as it elongates. The fact that $\alpha_P<0$ (resp. $\beta_R < 0$) confirms that in the absence of external forces these quantities relax back to the equilibrium shape $P=0$ (resp. $R=1$). Regarding the $x$-dependent external forces, $f_X$ exhibits a repulsive term $a_X/w^2$ that slows the cell near the entrance of the constriction, and an attractive term $b_X n_y/w^2$ that accelerates it near $x=0$, \emph{i.e.} once it is engaged in the constriction. The protrusion force $f_P$ exhibits a term $c_P n_x/w^2$ that is odd under reflection symmetry and drives the rapid change of sign of the protrusion $P$ as the nucleus crosses the tightest point of the constriction. Finally, the dynamics of $R$ is captured by a single, elongation-driving term $a_R/w$ with $a_R > 0$; the relaxation back to the equilibrium value at the constriction exit is accelerated by the coupling $\alpha_R P$ with negative $P$ values. All in all, this model thus recapitulates with a few terms the directed migration of the cells through the channel, and the way the nuclei stall when reaching the constriction entrance, then protrude, elongate, and finally pop through rapidly. In the final stage, the protrusion reverts and points backward, leading to a rapid relaxation of the aspect ratio and the exit from the constriction.

Our inference method also provides us with a physically interpretable estimate of the diffusion coefficients of the nucleus position $D_{X} \sim $\SI{5.8e-2}{\micro\meter^{2}\min^{-1}}. This value is several orders of magnitude above the equilibrium expectations from the Stokes-Einstein equation for a purely passive particle in the highly viscous cellular environment, $D_{\mathrm{Stokes-Einstein}} \sim $\SI{8e-7}{\micro\meter^{2}\min^{-1}}, reflecting the fact that cellular motion is activity-driven. Note that for simplicity of the analysis, we have assumed constant diffusion coefficients and Gaussian white noise. To investigate further these assumptions, larger amounts of data with a higher time resolution would be needed.

Finally, while the learned model provides good agreement in terms of capturing the dynamical geometric change of the nucleus during translocation, with a single parametric model encompassing the multiple constriction widths available, we note that it also presents some limitations. Indeed, this model is trained on a population of cells, and neglects any cell-to-cell variability due, \emph{e.g.}, to different sizes, genetic expression levels and age of the cells. This inherent variability manifests itself in a different way from the dynamical stochasticity captured here by the diffusion terms. Taking into account such cell-to-cell variability is a major challenge, as the amount of data available for each cell is small: data-efficient methods such as SFI~\cite{frishman_learning_2020} or Underdamped Langevin Inference~\cite{bruckner_inferring_2020} provide a promising avenue towards this, but single-event processes such as nuclear translocation studied here remain intractable with these approaches. A further difficulty comes from the limited frame rate, which leads to trajectories that appear to "tunnel through" right at the end of the passage through the constriction (as evidenced by long straight lines connecting data points in Fig~\ref{fig:Figure2}c) and lower the resolution of the translocation event. These challenges preclude the quantitative prediction of, \emph{e.g.}, mean translocation times, using the learned model.

\begin{figure*}[tb!]
\includegraphics[scale=0.3]{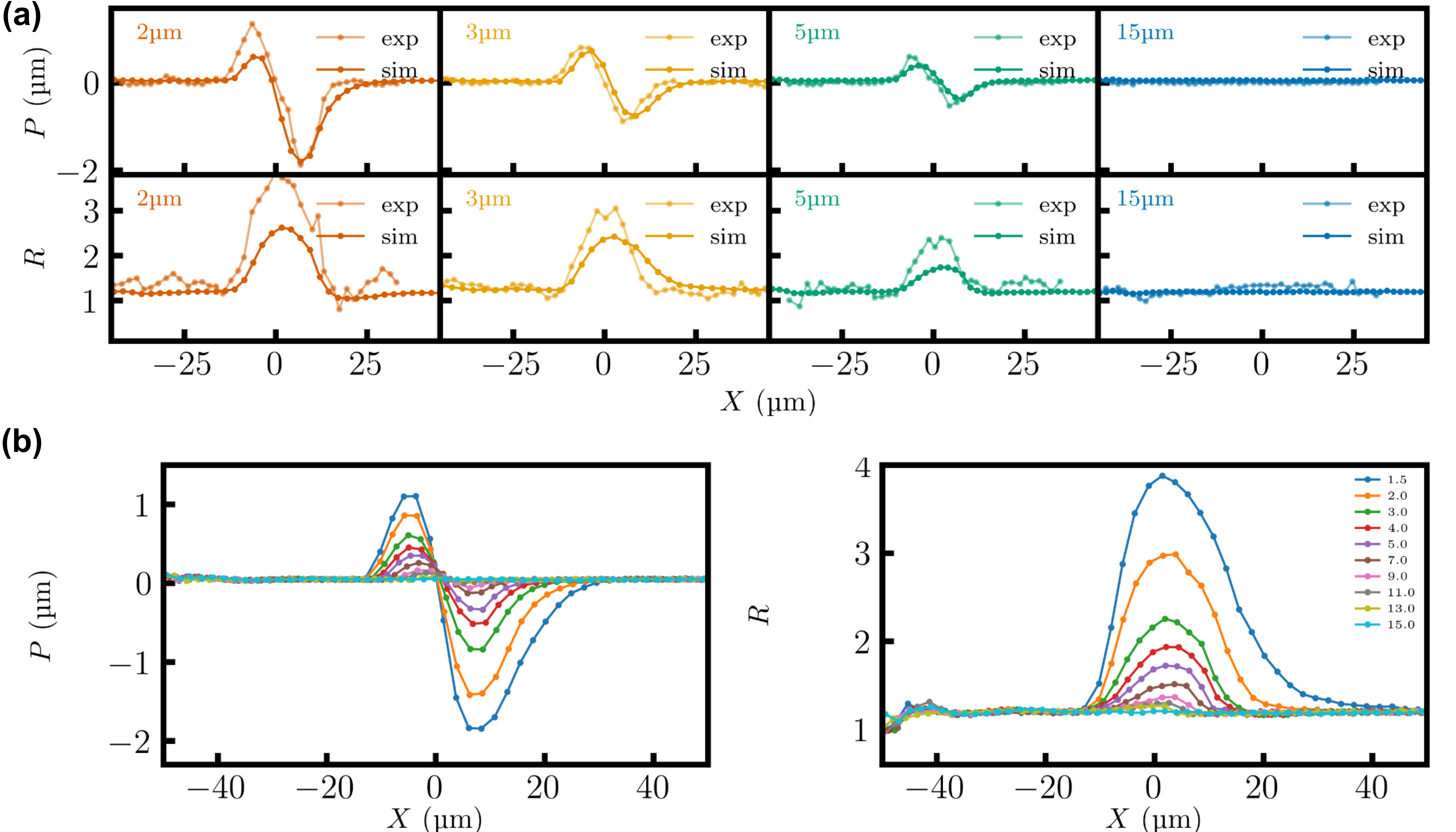}
\caption{\textbf{Predictivity of the inferred model to other geometries.} \textbf{(a)} Extrapolation from partial data compared with the experimental data. Comparison of the boundary polarity $P$ against nuclei position $X$ (first row) and aspect ratio against $X$ (second row). \textbf{(e, f)} Extrapolation over a range of constraints with widths listed in the legend.}
\label{fig:Figure5}
\end{figure*}

\paragraph{Predictivity of the model}
As our learned model takes the constriction geometry as an explicit parameter, we can extrapolate it to predict nuclear translocation dynamics in other geometries. To assess the validity of this approach, we first test it on geometries for which experimental results are available: to this aim, we perform again the inference while masking one of our four constriction sizes (2, 3, 5 and \SI{15}{\micro\meter}). We then use the model inferred from the other three constriction widths to make predictions on the fourth geometry, which includes interpolations (when masking the 3 and 5\unit{\micro\meter} sets) and extrapolations (when masking 2 and 15\unit{\micro\meter} data). We then compare in Fig.\ref{fig:Figure5}a the prediction on the fourth, masked geometry with the actual experimental results. The good general agreement confirms the validity of this approach, and shows the usefulness of the learned model to predict behavior in geometries not used to train it. Note, however, that the simulation results exhibit smoother geometrical deformation than the experiments, in particular past the constriction. This discrepancy could be due to the small number of data points during the short time interval when the nuclei pass through the constraint.

Next, we extrapolate the model to other geometries that were not studied experimentally: using the inferred model trained on the full data set, we simulate trajectories and compute the average geometric quantities $P(X)$ and $R(X)$ in constraints ranging from \SI{1.0}{\micro\meter} to \SI{15}{\micro\meter} (Fig.\ref{fig:Figure5}b). Each curve is obtained by averaging over \num{1000} simulated trajectories.  We observe a continuous increase in the maximum aspect ratio and geometric polarity as the constraint gets smaller. These predictions could be used for future experiment design, as a way to explore parameter space and focus experiments on the regions of interest.

\section{Discussion}

In this article, we have studied the spontaneous migration of individual cells in a microfluidic device that exerts tight three-dimensional constraints mimicking physiological scenarii where cells are able to migrate in strongly confined environments.  Strikingly, cells can pass through constrictions much smaller than the rest diameter of their nucleus, leading to large  deformations of the nucleus during translocation ~\cite{davidson_nuclear_2014,davidson_design_2015}. This controlled experimental setup differs from previous studies of 2D confined cell migration without three-dimensional constraints~\cite{bruckner_stochastic_2019}, in which the nucleus is not significantly deformed. We segment and track cell nuclei to obtain trajectories that we use to quantify the dynamics of this nuclear translocation process. To this aim, we employ a data-driven approach that captures the stochastic nature of the motion and shape changes of the nucleus during cell motility in strongly constraining environments. In contrast with previous works where only the nucleus position was used~\cite{bruckner_learning_2021}, leading to effectively inertial dynamics, we include shape descriptors in our model that provide a proxy for the unobserved polarity of the cell. The outcome is an optimized set of overdamped equations that quantitatively captures the joint dynamics of nuclear position, protrusion and elongation as coupled stochastic differential equations. Importantly, the geometry of the environment is an explicit parameter of the resulting model, which allows for predictions and extrapolation to other geometries. 

Nuclear translocation involves a complex set of molecular mechanisms that enables cells to sense their mechanical environment and adapt their internal forces. Our study paves the way towards a data-driven understanding of this process,  where the nucleus is considered as an actor of the dynamical process, rather than a passive tracer lagging behind. In the future, this description could be enriched with other cell state descriptors, in particular with the spatial distribution of cytoskeletal and nuclear components, such as protein complexes involved in the mechanotransduction process. A challenge towards this, however, consists in selecting appropriate quantitative descriptors to include in the dynamical model: for instance, while nesprin -- the mechanical linking protein between cytoskeleton and nucleus -- is observed to accumulate at the front of the nucleus, this is not recapitulated by a polarity defined in terms of the first moment of the protein distribution (see Appendix B3).

\section{Acknowledgments}
We thank Thierry Dubois and his team (Breast Cancer Biology Group, Institut Curie - PSL Research University, Paris, France), as well as Julie Plastino and Inge Bos (Active Cell Matter team, LPENS, Paris, France), Chase Broedersz (Vrije Universiteit, Amsterdam, The Netherlands), David Br\"uckner (Institute of Science and Technology Austria) and Jan Lammerding (Cornell University, USA) for their help, support and fruitful discussions. This work was carried out in part at the Nikon Imaging Centre, Institut Curie, member of the
French National Research Infrastructure France-BioImaging [ANR10-INBS-04]. The project leading to this publication was supported by an institutional Institut Curie grant and received funding from France 2030, the French Government program managed by the French National Research Agency (ANR-16-CONV-0001) and from Excellence Initiative of Aix-Marseille University - A*MIDEX.

\section{Appendix A : Material and Methods}\label{sec:MM}

\paragraph{A1. Cell Culture}
Mouse Embryonic Fibroblasts (MEF) were CRISPR-modified to create a new cell line: MEFs SYNE2-GFP LMNA-mCh as described and validated in \cite{davidson_nesprin2_2020}. Cells are cultured at 37°C in a humidified incubator with 5\% CO\textsubscript{2}, in DMEM (Dulbecco's Modified Eagle Medium - Gibco) supplemented with 10\% (v/v) Fetal Bovine Serum (FBS – Gibco). 

\paragraph{A2. Migration Devices}
The epoxy mold (R123/R614 - Soloplast) we used was replicated from a polydimethylsiloxane (PDMS) imprinted piece coming from the lab of Jan Lammerding (Cornell University, USA). A mix of PDMS (using a 10:1 ratio polymer:crosslinker) is vacuumed for 20 minutes to avoid bubbles, then poured into the epoxy mold and let to cure for 4 hours in a 60°C oven. Imprinted PDMS pieces are cut using a scalpel and biopsy punches (2mm and 5mm in diameter). Glass coverslips are soaked overnight in a 0.2M solution of HCl and rinsed with H\textsubscript{2}O and ethanol, dried with Kim wipes. To form a migration device, an imprinted PDMS piece and a treated glass coverlip are placed in a plasma cleaner for 1 minute and gently sticked together. This process creates covalent bonds between the PDMS and the glass \cite{borok_pdms_2021}. Devices are then directly put on a 100°C hot plate for 5 minutes to help the sticking process.

\paragraph{A3. Cell migration experiment}
Microfluidic devices are sterilized and rinsed under a microbiological safety post: first once with ethanol ($\sim$250\unit{\micro\liter}) then twice with Phosphate-buffered saline (PBS - Gibco) and twice with DMEM (Gibco) supplemented with 10\% (v/v) FBS. Cells are suspended at a concentration of 10 millions per mL in DMEM (Gibco) supplemented with 10\% (v/v) FBS. They are seeded in the device by adding 5\unit{\micro\liter} of the suspended solution in one of the two small ports of the device. After 6 hours, enough cells are in the constricted region of the device and acquisition can start. For that, cell medium is changed to DMEM without phenol red and with HEPES (15 mM) (Gibco), supplemented with 10\% FBS (Gibco), 100 units/mL penicillin, and 100 µg/mL streptomycin (Life Technologies).

\paragraph{A4. Image Acquisition}
Timelapse acquisitions are performed on an epifluorescence microscope (Nikon Ti-E) equipped with a sCMOS camera (2048 ORCA Flash 4.0 V2, Hamamatsu or Prime BSI, Teledyne), a perfect focus system, a 60x oil objective (Nikon), and a temperature and gas control chamber (set on 37°C, air at 5\% CO\textsubscript{2}). Images are taken every 10 minutes.

\paragraph{A5. Image Analysis}
Movies are analyzed using Image J/Fiji and Python. The projected nucleus surface is detected by using the "analyze particles" function on a threshold (median filter to 5.0 radius, normalized by 0.4\% and autolocal threshold "Bernsen" 5) applied on the mCherry image (corresponding to a lamin A/C signal). The nucleus contour is defined by a band of 1\unit{\micro\meter} width created from the detected nucleus projected surface ("reduce" and "make band" functions).

\paragraph{A6. Definition of the spatial origin}
The origin of the $x$ axis is set at the center of the constriction pillar (2, 3, \SI{5}{\micro\meter}) or half pillar (\SI{15}{\micro\meter}). The origin of the $y$ axis is set at the top center of the bottom constriction pillar (2, 3, \SI{5}{\micro\meter}) or the top center of the fitted circle to the bottom half pillar (\SI{15}{\micro\meter}).

\paragraph{A7. Definition of $X, X_c, P, R$}
The position of a nucleus is defined by its surface barycenter ($X, Y$),  specifically $X=\sum_{i=0}^{n}x_{i}/n$ and $Y=\sum_{i}y_i/n$ with $(x_{i},y_i)$ the coordinates of each pixel $i$ of the nucleus surface and $n$ the number of pixels in the nucleus surface. 
The $x$ coordinate of the center of the nucleus contour $X_c$, is defined as $X_c = \sum_{i=0}^{n_c}x_{c,i}/n_c$, with $x_{c,i}$ the $x$ coordinates of each pixel of the nucleus contour and $n_c$ the number of pixels in the nucleus contour.
The nucleus protrusion vector is defined as $P = X_{c} - X$.
The aspect ratio of the nucleus is defined by $R=R_{x}/R_{y}$ where $R_{x} = \sqrt{\sum_{i=0}^{n} (x_{i} - X)^{2}/n}$ and $R_{y} = \sqrt{\sum_{i=0}^{n} (y_{i} - Y)^{2}/n}$.

\begin{figure}[h!]
    \centering
    \includegraphics[width=0.45\textwidth]{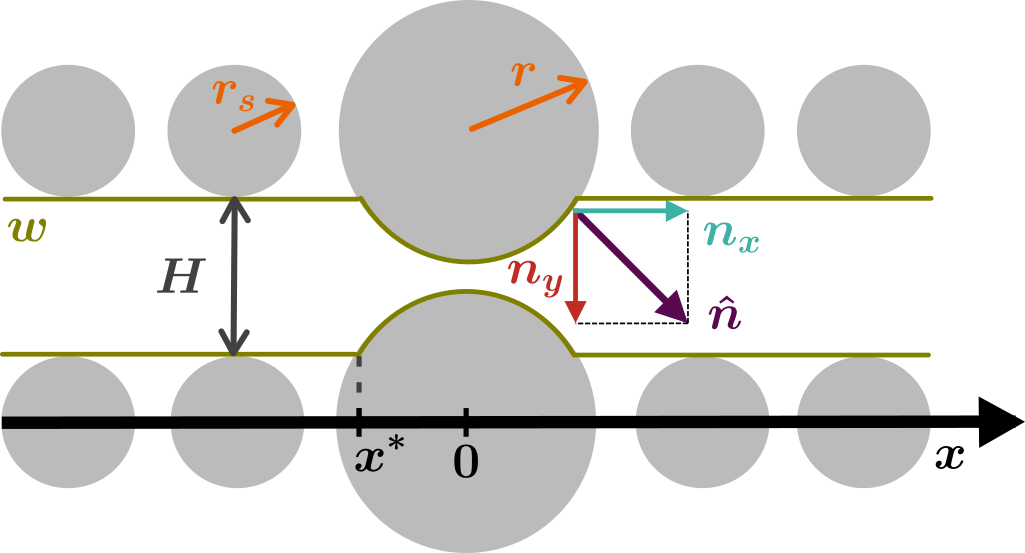}
    \caption{\textbf{Schematics of the geometric quantities describing the pillar shape used to construct force estimators.} Note that, for one series the centers of the pillars are aligned on $x$.}
    \label{fig:enter-label}
\end{figure}

\paragraph{A8. Complementary expressions for basis functions} 
The direct and indirect effects of the environment, $f_{ext}(X, r), f_{P}(X, r)$ and $f_{R}(X, r)$ are approximated by combining the width function of the channel $w(X,r)$ with the normal vector calculated from the shape of the pillar which is a circle of radius r. The normal vector ${\hat{n}}(X,r) = (n_{x}(X,r), n_{y}(X,r))$ of a circle of radius r is given by
\begin{align}
n_{x}(x,r) &= \dfrac{x}{r}, \qquad -x^{*} < x < x^{*}, \\
n_{x}(x,r) &= 0, \qquad \mathrm{otherwise} \\
n_{y}(x,r) &= \dfrac{\sqrt{r^{2} - x^{2}}}{r}, \qquad -x^{*} < x < x^{*}, \\
n_{y}(x,r) &= \dfrac{\sqrt{r^{2} - x^{*2}}}{r} \qquad \mathrm{otherwise}.
\end{align}
where $x^*=\sqrt{r^2-r_s^2}$ with $r_s$ the small pillar radius. These quantities are schematized in Fig.\ref{fig:enter-label}. 
The channel width $w(x)$ is given by
\begin{align}
w(x,r) = H+2r_{\mathrm{s}} - 2 \sqrt{r^{2} - x^{2}} &\qquad -x^{*} < x < x^{*}, \\
w(x,r) = H &\qquad \mathrm{otherwise},  
\end{align}
where $H=15$ is the channel height (note that we neglect the texture of the small pillars here, as they do not constrict the nucleus). 

\paragraph{A9. Full expression of the initial model}
The full model consisting of all basis functions, constructed by systematic expansion of the model over physically relevant variables, consists in linear combinations of the following basis functions:
\begin{multline}
        \{ 1, P, (R-R^{-1}), \\
        1/w(X,r),n_x(X,r)/w(X,r),n_y(X,r)/w(X,r),\\ 1/w^2(X,r),n_x(X,r)/w^2(X,r),n_y(X,r)/w^2(X,r), \\1/w^3(X,r),n_x(X,r)/w^3(X,r),n_y(X,r)/w^3(X,r)\}
\end{multline}

and has 39 parameters ($3\times 12$ for the drift, $3$ for the diffusion). The model inference and reduction framework described in Fig~\ref{fig:workflow} retains only 14 significant terms.

\paragraph{A10. Improvement of the SFI algorithm for large time intervals $\Delta t$}
To learn the dynamics of our system, characterized in this paragraph by the vector $\mathbf{x}_t\equiv (X_t,P_t,R_t)$, it is essential to estimate its discrete time derivative using $\Delta \mathbf{x}_{t} = \mathbf{x}_{t + \Delta t}- \mathbf{x}_{t}$. A challenge on applying SFI on cellular dynamics data is that the time interval $\Delta t$ between frames is large compared (10 minutes) and cannot be easily reduced due to phototoxicity: with the previously introduced algorithm~\cite{frishman_learning_2020}, this incurs $O(\Delta t)$ biases on the estimators. To adapt the method to this challenge, we propose a modification, which results in much smaller $O(\Delta t^2)$ biases.

Specifically, we focus in the derivation on approximate time differences $\Delta \mathbf{x}_t$ rather than the infinitesimal time difference $d \mathbf{x}_t$. Writing the dynamics in a generic form $\frac{\mathrm{d}\mathbf{x}}{\mathrm{d}t} = \mathbf{f}(\mathbf{x}_t) + \sqrt{2\mathbf{D}}\eta_t$, we have for discrete time increments:
\begin{align} %XXX CHECK
    \frac{\Delta \mathbf{x}_t}{\Delta t} &= \frac{1}{\Delta t} \int_t^{t+\Delta t} \frac{d\mathbf{x}_{t'}}{dt} dt' \\
    &=  \frac{1}{\Delta t} \int_t^{t+\Delta t}  \left[ \mathbf{f}(\mathbf{x}_{t'}) + \sqrt{2\mathbf{D}}\eta_{t'} \right] \mathrm{d}t'
\end{align}
SFI consists in approximating the unknown deterministic drift field $\mathbf{f}(\mathbf{x})$ by a linear combination of basis functions $\mathbf{f}(\mathbf{x}) = \sum_{\alpha} F_{\alpha} \mathbf{b}_{\alpha}(\mathbf{x})$ where $F_{\alpha}$ are the coefficients to infer and $\mathbf{b}_{\alpha}(\mathbf{x})$ are the basis functions. Thus, we can project the above equation on one of the basis functions $\mathbf{b}_{\gamma}(\mathbf{x}_t)$ and derive its average in the Itô convention:
\begin{multline}
    \left \langle \frac{\Delta  \mathbf{x}_t}{\Delta t} \mathbf{b}_{\gamma}(\mathbf{x}_t) \right \rangle =  \sum_{\alpha}  F_{\alpha} \left \langle b_{\gamma}(\mathbf{x}_t) \frac{1}{\Delta t} \int_{t}^{t+\Delta t}\!\! b_{\alpha}(\mathbf{x}_{t'})  dt' \right \rangle
    \label{eq:projected_dynamic}
\end{multline}
where $ \langle \cdot\rangle$ represents the expectation over many realisations of the noise $\eta_t$, conditioned on the initial value $\mathbf{x}_t$. Since we only measure $\mathbf{x}$ at discrete times $t, t+\Delta t, \dots$, we need to approximate $\int_{t}^{t+\Delta t} b_{\alpha}(\mathbf{x}_{t'})  dt' \approx \frac{\Delta t}{2}(b_{\alpha}(\mathbf{x}_{t}) + b_{\alpha}(\mathbf{x}_{t + \Delta t}))$. Importantly, this \emph{trapezoidal integration rule} is a more accurate approximation than the Riemann sum approximation $b_{\alpha}(\mathbf{x}_{t})\Delta t$ used in Ref.~\cite{frishman_learning_2020}.

By now averaging Eq.\ref{eq:projected_dynamic} over all data points $\{\mathbf{x}_{t_i}\}_{i=1,N}$ and inverting the \emph{trapezoidal} approximation for the matrix $\sum_{i=1}^{N-1} \left(\frac{1}{\Delta t} \int_{t}^{t+\Delta t} b_{\alpha}(\mathbf{x}_{t'})  dt'\right)\mathbf{b}_{\gamma}(\mathbf{x}_{t_i})$, we derive a corrected estimator of $F_\alpha$ for large $\Delta t$:
\begin{multline}
    \hat{F}_{\alpha} = \sum_{\gamma} \left(\sum_{i=1}^{N-1} \frac{1}{2} \left(\mathbf{b}_{\alpha}(\mathbf{x}_{t_i}) + \mathbf{b}_{\alpha}(\mathbf{x}_{t_i + \Delta t})\right)\mathbf{b}_{\gamma}(\mathbf{x}_{t_i}) \right)^{-1}\\
    \sum_{i=1}^{N-1} \frac{\Delta  \mathbf{x}_{t_i}}{\Delta t} \mathbf{b}_{\gamma}(\mathbf{x}_{t_i}).
\end{multline}
The use of the trapezoidal method for discrete differences thus results in a lower-order discretization bias compared to the original SFI method, and enables accurate inference with the data set considered in this article.

\paragraph{A11. Bootstrap methods for coefficient error estimation}
We estimate the mean and standard deviation of each coefficient and the diffusion constant using the bootstrap method. More specifically, we sample with replacements the set of small 5-consecutive-points trajectories to generate an ensemble of trajectories~\cite{tibshirani_introduction_1994}. For each sample, the drift coefficients are estimated with the above-mentioned modified SFI algorithm, while the diffusion coefficients are estimated using the method introduced by Vestergaard \emph{et al.}~\cite{vestergaard_optimal_2014,frishman_learning_2020}. 
We compute the average and standard deviation of coefficients over 20 bootstrapped data sets obtained from the initial set of trajectories, and use the resulting standard deviations as indicators of the confidence interval for our assessment of the statistical significance of these coefficients.

%------------------------------------------------------------------%

\section{Appendix B : Complementary results}

\begin{center}
\begin{figure}[hbt]
    \includegraphics[]{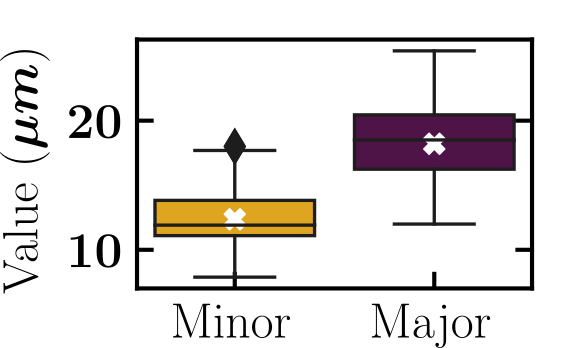}
    \caption{\textbf{MEFs nuclei size is measured through minor and major axis of a fitted ellipse.} Width (\textit{minor axis of the fitted ellipse}) and length (\textit{major axis}) of MEFs nuclei measured inside the microfluidic device when $X=-20 \pm5$\unit{\micro\meter}, regardless of the constriction size. White cross represents the average value of the distribution. N = 75} 
    \label{app:SizeNuclei}
\end{figure}
\end{center}

\paragraph{B1. Size of MEFs nuclei}
The apparent size of MEFs nuclei in the microfluidic device is calculated at $X=-20$\unit{\micro\meter}, before they are deformed by the constrictions. In fact, we allow a confidence interval of 5\unit{\micro\meter}. We fit an ellipse to the detected nucleus and measure its major and minor axis (Fig.\ref{app:SizeNuclei}).

\begin{center}
\begin{figure}[hbt]
    \includegraphics[width=0.48\textwidth]{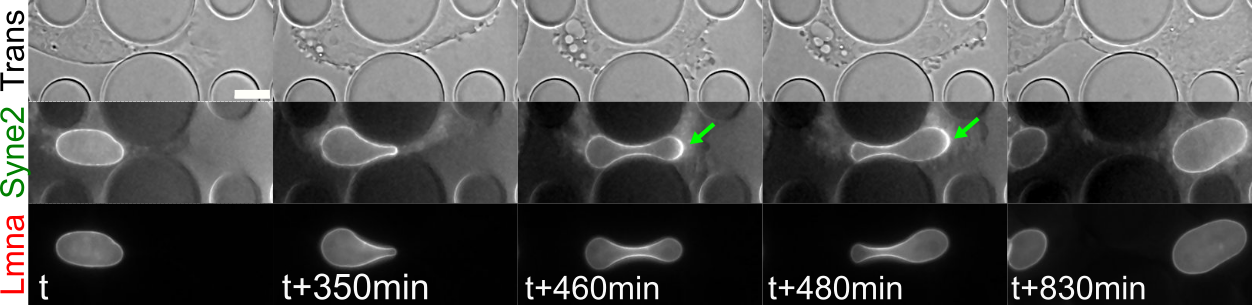}
    \caption{\textbf{MEF cell going through a \SI{2}{\micro\meter} wide constriction.} \textit{Top}: Transmission signal. \textit{Middle} : GFP signal corresponding to the expression of Syne2-GFP gene for nesprin 2G proteins. Arrows point at hyperfluorescence of nesprin at the front of the nucleus during translocation. \textit{Bottom}: mCherry signal corresponding to the expression Lmna-mCh gene for lamin A/C proteins. Scale bar : \SI{10}{\micro\meter}.} 
    \label{app:Timelapse}
\end{figure}
\end{center}

\paragraph{B2. Timelapse of MEFs going through a \SI{2}{\micro\meter} constriction}
In agreement with previous results \cite{davidson_nesprin2_2020}, we observe that when cells migrate through \SI{2}{\micro\meter} constrictions (Fig.\ref{app:Timelapse}, \textit{Trans}), nesprin accumulates at the front of the nucleus (Fig.\ref{app:Timelapse}, \textit{Syne 2 -- green arrows}) whereas lamin do not (Fig.\ref{app:Timelapse}, \textit{Lmna}).

\begin{figure}[hbt]
        \includegraphics[width=0.45\textwidth]{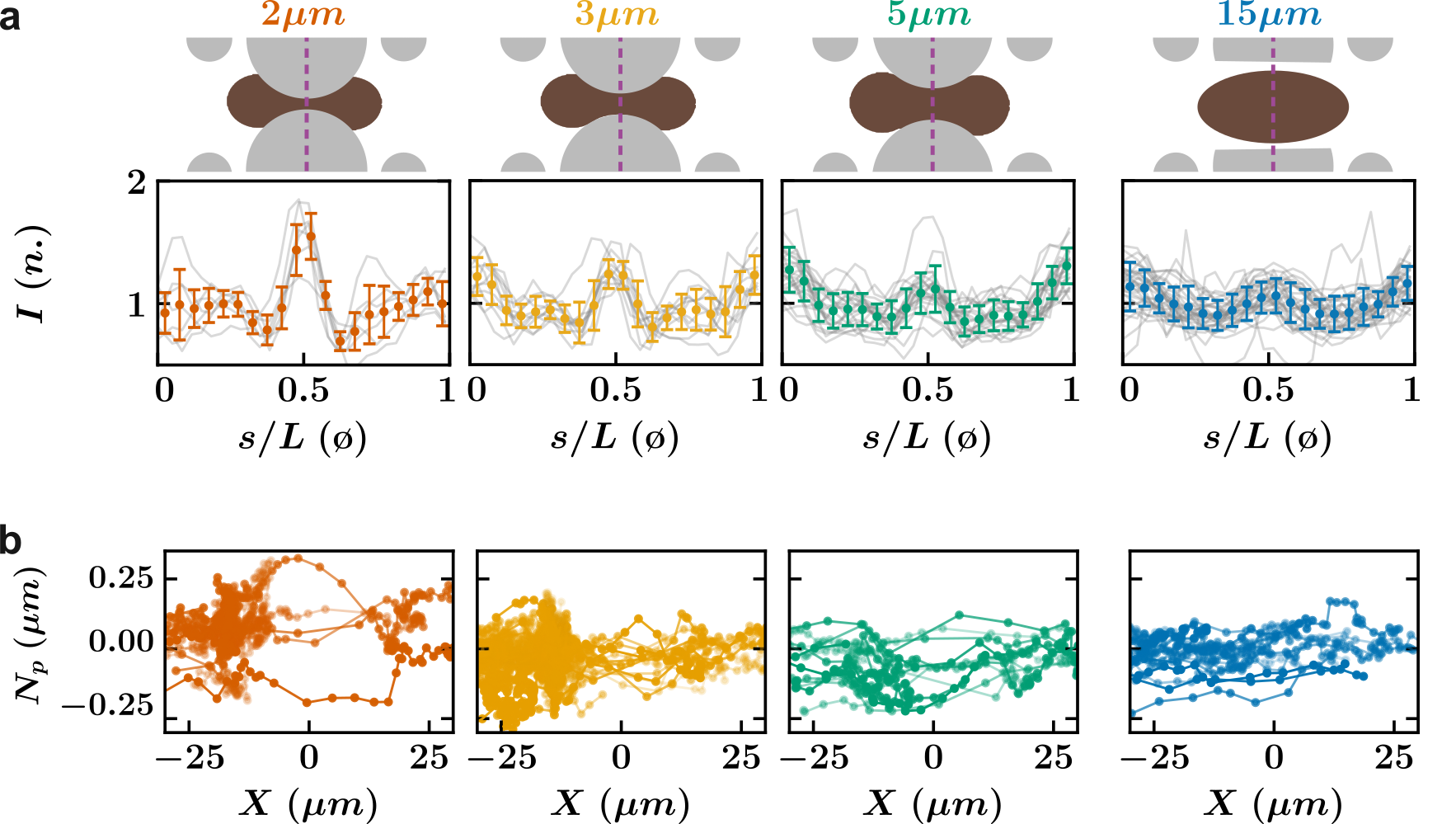}
        \caption{\textbf{Nesprins accumulate at the front of their nuclei but the defined polarity vector does not show this tendency.} \textbf{a}.Nesprin 2G intensity signals measured along the curvilinear abscissa $s$ when the nucleus (of perimeter $L$) is in the middle of the constriction (resp. large channel) : $X=\SI{0\pm5}{\micro\meter}$ Intensities are normalised by the contrast and divided by the average on the contour. From left to right number of cells shown is: 7, 11, 23 and 32. Error bars correspond to standard deviation. \textbf{b}. Calculated nesprin polarity vector ($N_p$) according to nucleus position ($X$) for different constrictions sizes (2, 3 and \SI{5}{\micro\meter}) as well as large channels (\SI{15}{\micro\meter}). Each line represents a single cell. Dots are experimental data points. From left to right  number of cells shown is: 14, 36, 15 and 20.} 
        \label{app:NesprinVector}
\end{figure}

\paragraph{B3. Nesprin distribution polarity vector}
For each constriction size (2, 3 and \SI{5}{\micro\meter}) and for the \SI{15}{\micro\meter} large channel, when the nucleus is in the middle of the constriction, we  measure the intensity along a curvilinear abscissa ($s$) around the nucleus (of perimeter $L$) at $X=\SI{0}{\micro\meter}$ (Fig.\ref{app:NesprinVector}a). At $s/L=0.5$, which corresponds to the front of the nucleus, we confirm the increase of the nesprin signal intensity for the three constriction sizes (2, 3 and \SI{5}{\micro\meter}) that does not happen in the \SI{15}{\micro\meter} large channel. This accumulation of nesprin at the front of the nucleus appears to be more dramatic as the constriction gets smaller. 

In order to account for the nesprin distribution in a quantitative unidimensional way, we construct a \emph{nesprin polarity vector}. First, we define $X_n$ as the barycenter of the nucleus contour weighted by nesprin intensity. Explicitly, 
\begin{equation}
    X_n = \frac{\sum_{i=0}^{n_c}x_{c,i} I_{c,i}}{\sum_{i=0}^{n_c}I_{c,i}}
\end{equation} 
with ($I_{c,i}$) the corresponding intensity values of each pixels ($x_{c,i}$) of the nucleus contour.
Second, we define a nesprin polarity vector $N_p$ as $N_p = X_n - X_c$. It is calculated for cells migrating through small constrictions (2, 3 and 5 \unit{\micro\meter}) and large channels (15\unit{\micro\meter}) along nucleus position $X$ (Fig.\ref{app:NesprinVector}.b). In fact, $N_p$ is noisy, almost constant over nucleus position and close to zero. In addition, there is no significant difference between the nesprin polarity of cells migrating through small constrictions and of cells migrating through large channels, when experimentally nesprins do not accumulate at the front of the nucleus in the 15\unit{\micro\meter} case. This means that the barycenter of the contour weighted by the nesprin intensity $X_n$ and the barycenter of the contour $X_c$ are close in value. Therefore, the application of our method to dynamical changes in protein distribution will need further investigation, in particular to account for nesprin accumulation at the nucleus front.

\bibliographystyle{unsrt}
\bibliography{SqueezedCellsInference.bib}

\begin{thebibliography}{10}

\bibitem{guilak_viscoelastic_2000}
Farshid Guilak, John~R. Tedrow, and Rainer Burgkart.
\newblock Viscoelastic {Properties} of the {Cell} {Nucleus}.
\newblock {\em Biochemical and Biophysical Research Communications},
  269(3):781--786, 2000.

\bibitem{uhler_regulation_2017}
Caroline Uhler and G.~V. Shivashankar.
\newblock Regulation of genome organization and gene expression by nuclear
  mechanotransduction.
\newblock {\em Nat Rev Mol Cell Biol}, 18(12):717--727, 2017.

\bibitem{davidson_nuclear_2014}
Patricia~M. Davidson, Celine Denais, Maya~C. Bakshi, and Jan Lammerding.
\newblock Nuclear {Deformability} {Constitutes} a {Rate}-{Limiting} {Step}
  {During} {Cell} {Migration} in 3-{D} {Environments}.
\newblock {\em Cellular and Molecular Bioengineering}, 7(3):293--306, 2014.

\bibitem{wolf_physical_2013}
Katarina Wolf, Mariska Te~Lindert, Marina Krause, Stephanie Alexander, Joost
  Te~Riet, Amanda~L. Willis, Robert~M. Hoffman, Carl~G. Figdor, Stephen~J.
  Weiss, and Peter Friedl.
\newblock Physical limits of cell migration: {Control} by {ECM} space and
  nuclear deformation and tuning by proteolysis and traction force.
\newblock {\em Journal of Cell Biology}, 201(7):1069--1084, 2013.

\bibitem{patteson_vimentin_2019}
Alison~E. Patteson, Amir Vahabikashi, Katarzyna Pogoda, Stephen~A. Adam,
  Kalpana Mandal, Mark Kittisopikul, Suganya Sivagurunathan, Anne Goldman,
  Robert~D. Goldman, and Paul~A. Janmey.
\newblock Vimentin protects cells against nuclear rupture and {DNA} damage
  during migration.
\newblock {\em Journal of Cell Biology}, 218(12):4079--4092, 2019.

\bibitem{estabrook_calculation_2021}
Ian~D. Estabrook, Hawa~Racine Thiam, Matthieu Piel, and Rhoda~J. Hawkins.
\newblock Calculation of the force field required for nucleus deformation
  during cell migration through constrictions.
\newblock {\em PLOS Computational Biology}, 17(5):e1008592, 2021.

\bibitem{verkhovsky_self-polarization_1999}
Alexander~B. Verkhovsky, Tatyana~M. Svitkina, and Gary~G. Borisy.
\newblock Self-polarization and directional motility of cytoplasm.
\newblock {\em Current Biology}, 9(1):11--S1, 1999.

\bibitem{maiuri_actin_2015}
Paolo Maiuri, Jean-François Rupprecht, Stefan Wieser, Verena Ruprecht, Olivier
  Bénichou, Nicolas Carpi, Mathieu Coppey, Simon De Beco, Nir Gov,
  Carl-Philipp Heisenberg, Carolina Lage Crespo, Franziska Lautenschlaeger,
  Maël Le Berre, Ana-Maria Lennon-Dumenil, Matthew Raab, Hawa-Racine Thiam,
  Matthieu Piel, Michael Sixt, and Raphaël Voituriez.
\newblock Actin {Flows} {Mediate} a {Universal} {Coupling} between {Cell}
  {Speed} and {Cell} {Persistence}.
\newblock {\em Cell}, 161(2):374--386, 2015.

\bibitem{keren_mechanism_2008}
Kinneret Keren, Zachary Pincus, Greg~M. Allen, Erin~L. Barnhart, Gerard
  Marriott, Alex Mogilner, and Julie~A. Theriot.
\newblock Mechanism of shape determination in motile cells.
\newblock {\em Nature}, 453(7194):475--480, 2008.

\bibitem{jaalouk_mechanotransduction_2009}
Diana~E. Jaalouk and Jan Lammerding.
\newblock Mechanotransduction gone awry.
\newblock {\em Nat Rev Mol Cell Biol}, 10(1):63--73, 2009.

\bibitem{maurer_driving_2019}
Melanie Maurer and Jan Lammerding.
\newblock The {Driving} {Force}: {Nuclear} {Mechanotransduction} in {Cellular}
  {Function}, {Fate}, and {Disease}.
\newblock {\em Annu. Rev. Biomed. Eng.}, 21(1):443--468, 2019.

\bibitem{venturini_nucleus_2020}
Valeria Venturini, Fabio Pezzano, Frederic Català~Castro, Hanna-Maria
  Häkkinen, Senda Jiménez-Delgado, Mariona Colomer-Rosell, Monica Marro,
  Queralt Tolosa-Ramon, Sonia Paz-López, Miguel~A. Valverde, Julian Weghuber,
  Pablo Loza-Alvarez, Michael Krieg, Stefan Wieser, and Verena Ruprecht.
\newblock The nucleus measures shape changes for cellular proprioception to
  control dynamic cell behavior.
\newblock {\em Science}, 370(6514):eaba2644, 2020.

\bibitem{lomakin_nucleus_2020}
A.~J. Lomakin, C.~J. Cattin, D.~Cuvelier, Z.~Alraies, M.~Molina, G.~P.~F.
  Nader, N.~Srivastava, P.~J. Sáez, J.~M. Garcia-Arcos, I.~Y. Zhitnyak,
  A.~Bhargava, M.~K. Driscoll, E.~S. Welf, R.~Fiolka, R.~J. Petrie, N.~S.
  De~Silva, J.~M. González-Granado, N.~Manel, A.~M. Lennon-Duménil, D.~J.
  Müller, and M.~Piel.
\newblock The nucleus acts as a ruler tailoring cell responses to spatial
  constraints.
\newblock {\em Science}, 370(6514):eaba2894, 2020.

\bibitem{davidson_design_2015}
Patricia~M. Davidson, Josiah Sliz, Philipp Isermann, Celine Denais, and Jan
  Lammerding.
\newblock Design of a microfluidic device to quantify dynamic intra-nuclear
  deformation during cell migration through confining environments.
\newblock {\em Integr. Biol.}, 7(12):1534--1546, 2015.

\bibitem{gundersen_assembly_2018}
Jeremy Keys, Aaron Windsor, and Jan Lammerding.
\newblock Assembly and {Use} of a {Microfluidic} {Device} to {Study} {Cell}
  {Migration} in {Confined} {Environments}.
\newblock In Gregg~G. Gundersen and Howard~J. Worman, editors, {\em The {LINC}
  {Complex}}, volume 1840, pages 101--118. Springer New York, New York, NY,
  2018.

\bibitem{brunton_discovering_2016}
Steven~L. Brunton, Joshua~L. Proctor, and J.~Nathan Kutz.
\newblock Discovering governing equations from data by sparse identification of
  nonlinear dynamical systems.
\newblock {\em PNAS}, 113(15):3932--3937, 2016.

\bibitem{champion_data-driven_2019}
Kathleen Champion, Bethany Lusch, J.~Nathan Kutz, and Steven~L. Brunton.
\newblock Data-driven discovery of coordinates and governing equations.
\newblock {\em Proceedings of the National Academy of Sciences},
  116(45):22445--22451, 2019.

\bibitem{romeo_learning_2021}
Nicolas Romeo, Alasdair Hastewell, Alexander Mietke, and Jörn Dunkel.
\newblock Learning developmental mode dynamics from single-cell trajectories.
\newblock {\em eLife}, 10:e68679, 2021.

\bibitem{schmitt_zyxin_2023}
Matthew~S. Schmitt, Jonathan Colen, Stefano Sala, John Devany, Shailaja
  Seetharaman, Margaret~L. Gardel, Patrick~W. Oakes, and Vincenzo Vitelli.
\newblock Zyxin is all you need: machine learning adherent cell mechanics,
  2023.

\bibitem{kloeden_numerical_2010}
Peter~E. Kloeden and Eckhard Platen.
\newblock {\em Numerical solution of stochastic differential equations}.
\newblock Number~23 in Applications of mathematics. Springer, Berlin, corr. 3.
  print edition, 2010.

\bibitem{selmeczi_cell_2005}
David Selmeczi, Stephan Mosler, Peter~H. Hagedorn, Niels~B. Larsen, and Henrik
  Flyvbjerg.
\newblock Cell {Motility} as {Persistent} {Random} {Motion}: {Theories} from
  {Experiments}.
\newblock {\em Biophysical Journal}, 89(2):912--931, 2005.

\bibitem{selmeczi_cell_2008}
D.~Selmeczi, L.~Li, L.~I.I. Pedersen, S.~F. Nrrelykke, P.~H. Hagedorn,
  S.~Mosler, N.~B. Larsen, E.~C. Cox, and H.~Flyvbjerg.
\newblock Cell motility as random motion: {A} review: {Cell} motility as random
  motion.
\newblock {\em Eur. Phys. J. Spec. Top.}, 157(1):1--15, 2008.

\bibitem{li_dicty_2011}
Liang Li, Edward~C. Cox, and Henrik Flyvbjerg.
\newblock '{Dicty} dynamics': {Dictyostelium} motility as persistent random
  motion.
\newblock {\em Phys Biol}, 8(4):046006, 2011.

\bibitem{bruckner_learning_2021}
David~B. Brückner, Nicolas Arlt, Alexandra Fink, Pierre Ronceray, Joachim~O.
  Rädler, and Chase~P. Broedersz.
\newblock Learning the dynamics of cell–cell interactions in confined cell
  migration.
\newblock {\em Proc. Natl. Acad. Sci. U.S.A.}, 118(7):e2016602118, 2021.

\bibitem{bruckner_geometry_2022}
David~B. Brückner, Matthew Schmitt, Alexandra Fink, Georg Ladurner, Johannes
  Flommersfeld, Nicolas Arlt, Edouard Hannezo, Joachim~O. Rädler, and Chase~P.
  Broedersz.
\newblock Geometry {Adaptation} of {Protrusion} and {Polarity} {Dynamics} in
  {Confined} {Cell} {Migration}.
\newblock {\em Phys. Rev. X}, 12(3):031041, 2022.

\bibitem{stoberl_nuclear_2023}
Stefan Stöberl, Johannes Flommersfeld, Maximilian~M. Kreft, Martin Benoit,
  Chase~P. Broedersz, and Joachim~O. Rädler.
\newblock Nuclear deformation and dynamics of migrating cells in {3D}
  confinement reveal adaptation of pulling and pushing forces, 2023.

\bibitem{frishman_learning_2020}
Anna Frishman and Pierre Ronceray.
\newblock Learning {Force} {Fields} from {Stochastic} {Trajectories}.
\newblock {\em Phys. Rev. X}, 10(2):021009, 2020.

\bibitem{bruckner_inferring_2020}
David~B. Brückner, Pierre Ronceray, and Chase~P. Broedersz.
\newblock Inferring the {Dynamics} of {Underdamped} {Stochastic} {Systems}.
\newblock {\em Phys. Rev. Lett.}, 125(5):058103, 2020.

\bibitem{davidson_nesprin2_2020}
Patricia~M Davidson, Aude Battistella, Théophile Déjardin, Timo Betz, Julie
  Plastino, Nicolas Borghi, Bruno Cadot, and Cécile Sykes.
\newblock Nesprin‐2 accumulates at the front of the nucleus during confined
  cell migration.
\newblock {\em EMBO Reports}, 21(7), 2020.

\bibitem{vahabikashi_nuclear_2022}
Amir Vahabikashi, Stephen~A. Adam, Ohad Medalia, and Robert~D. Goldman.
\newblock Nuclear lamins: {Structure} and function in mechanobiology.
\newblock {\em APL Bioengineering}, 6(1):011503, 2022.

\bibitem{gruenbaum_lamins_2015}
Yosef Gruenbaum and Roland Foisner.
\newblock Lamins: {Nuclear} {Intermediate} {Filament} {Proteins} with
  {Fundamental} {Functions} in {Nuclear} {Mechanics} and {Genome} {Regulation}.
\newblock {\em Annu. Rev. Biochem.}, 84(1):131--164, 2015.

\bibitem{tapley_connecting_2013}
Erin~C Tapley and Daniel~A Starr.
\newblock Connecting the nucleus to the cytoskeleton by {SUN}–{KASH} bridges
  across the nuclear envelope.
\newblock {\em Current Opinion in Cell Biology}, 25(1):57--62, 2013.

\bibitem{crisp_coupling_2006}
Melissa Crisp, Qian Liu, Kyle Roux, J.B. Rattner, Catherine Shanahan, Brian
  Burke, Phillip~D. Stahl, and Didier Hodzic.
\newblock Coupling of the nucleus and cytoplasm: {Role} of the {LINC} complex.
\newblock {\em The Journal of Cell Biology}, 172(1):41--53, 2006.

\bibitem{mellad_nesprins_2011}
Jason~A Mellad, Derek~T Warren, and Catherine~M Shanahan.
\newblock Nesprins {LINC} the nucleus and cytoskeleton.
\newblock {\em Current Opinion in Cell Biology}, 23(1):47--54, 2011.

\bibitem{denais_nuclear_2016}
Celine~M. Denais, Rachel~M. Gilbert, Philipp Isermann, Alexandra~L. McGregor,
  Mariska Te~Lindert, Bettina Weigelin, Patricia~M. Davidson, Peter Friedl,
  Katarina Wolf, and Jan Lammerding.
\newblock Nuclear envelope rupture and repair during cancer cell migration.
\newblock {\em Science}, 352(6283):353--358, 2016.

\bibitem{pfeifer_gaussian_2022}
Charlotte~R. Pfeifer, Michael~P. Tobin, Sangkyun Cho, Manasvita Vashisth,
  Lawrence~J. Dooling, Lizeth~Lopez Vazquez, Emma~G. Ricci-De~Lucca, Keiann~T.
  Simon, and Dennis~E. Discher.
\newblock Gaussian curvature dilutes the nuclear lamina, favoring nuclear
  rupture, especially at high strain rate.
\newblock {\em Nucleus}, 13(1):130--144, 2022.

\bibitem{bruckner_learning_2023}
David~B. Brückner and Chase~P. Broedersz.
\newblock Learning dynamical models of single and collective cell migration: a
  review, 2023.

\bibitem{drubin_origins_1996}
David~G Drubin and W.James Nelson.
\newblock Origins of {Cell} {Polarity}.
\newblock {\em Cell}, 84(3):335--344, 1996.

\bibitem{crutchfield_equations_1987}
James~P. Crutchfield and B.~S. McNamara.
\newblock Equations of motion from a data series.
\newblock {\em Complex systems}, 1:417--452, 1987.

\bibitem{bruckner_stochastic_2019}
David~B. Brückner, Alexandra Fink, Christoph Schreiber, Peter J.~F.
  Röttgermann, Joachim~O. Rädler, and Chase~P. Broedersz.
\newblock Stochastic nonlinear dynamics of confined cell migration in two-state
  systems.
\newblock {\em Nature Physics}, 15(6):595, 2019.

\bibitem{boninsegna_sparse_2018}
Lorenzo Boninsegna, Feliks Nüske, and Cecilia Clementi.
\newblock Sparse learning of stochastic dynamical equations.
\newblock {\em J. Chem. Phys.}, 148(24):241723, 2018.

\bibitem{callaham_nonlinear_2021}
J.~L. Callaham, J.-C. Loiseau, G.~Rigas, and S.~L. Brunton.
\newblock Nonlinear stochastic modelling with {Langevin} regression.
\newblock {\em Proceedings of the Royal Society A: Mathematical, Physical and
  Engineering Sciences}, 477(2250):20210092, 2021.

\bibitem{tibshirani_introduction_1994}
Bradley~Efron Tibshirani, R.~J.
\newblock {\em An {Introduction} to the {Bootstrap}}.
\newblock Chapman and Hall/CRC, New York, 1994.

\bibitem{borok_pdms_2021}
Alexandra Borók, Kristóf Laboda, and Attila Bonyár.
\newblock {PDMS} {Bonding} {Technologies} for {Microfluidic} {Applications}:
  {A} {Review}.
\newblock {\em Biosensors}, 11(8):292, 2021.

\bibitem{vestergaard_optimal_2014}
Christian~L. Vestergaard, Paul~C. Blainey, and Henrik Flyvbjerg.
\newblock Optimal estimation of diffusion coefficients from single-particle
  trajectories.
\newblock {\em Phys. Rev. E}, 89(2):022726, 2014.

\end{thebibliography}
    
\end{document}